\documentclass[a4paper]{jpconf}
\usepackage{graphicx}
\usepackage{wrapfig}
\begin{document}
\title{Longitudinal spin transfer of $\Lambda$ and $\bar\Lambda$ in polarized $pp$ collisions at $\sqrt s$=200 GeV at STAR}

\author{Qinghua Xu, for the STAR Collaboration}

\address{School of Physics, Shandong University, Shandong 250100, China}

\ead{xuqh@sdu.edu.cn}

\begin{abstract}
  We report our measurement on longitudinal spin transfer, $D_{LL}$, from high energy polarized protons to $\Lambda$ and $\bar{\Lambda}$ hyperons in proton-proton collisions at $\sqrt{s} = 200\,\mathrm{GeV}$ with the STAR detector at RHIC.
The current measurements cover $\Lambda$, $\bar\Lambda$ pseudorapidity $\left|\eta\right| < 1.2$ and transverse momenta $p_\mathrm{T}$ up to $4\,\mathrm{GeV}/c$ using the data taken in 2005.
The longitudinal spin transfer is found to be $D_{LL}= -0.03\pm 0.13(\mathrm{stat}) \pm 0.04(\mathrm{syst})$
 for inclusive $\Lambda$ and $D_{LL} = -0.12 \pm 0.08(\mathrm{stat}) \pm 0.03(\mathrm{syst})$ for inclusive $\bar{\Lambda}$ hyperons with  $\left<\eta\right> = 0.5$ and $\left<p_\mathrm{T}\right> = 3.7\,\mathrm{GeV}/c$.
The prospects with 2009 data and the future measurements are also given.
\end{abstract}

\section{Introduction}

$\Lambda$ hyperons have been studied extensively in different aspects of spin physics, due to their self-analyzing weak decay.
The longitudinal polarization of $\Lambda$ hyperons has been measured in $e^+e^-$ annihilation and lepton-nucleon deep inelastic scattering (DIS) with polarized beams and/or targets \cite{aleph}.
These studies can provide access to polarized fragmentation functions and the spin content of the $\Lambda$.
Here we report our measurement on longitudinal spin transfer ($D_{LL}$) from the proton beam to $\Lambda ({\bar \Lambda})$ produced in proton-proton collisions at $\sqrt s$=200 GeV\cite{STAR_DLL},
\begin{equation}
D_{LL}\equiv \frac
{\sigma_{p^+p \to  \Lambda ^+ X}-\sigma_{p^+p \to  \Lambda ^-X}}
{\sigma_{p^+p \to  \Lambda ^+ X}+\sigma_{p^+p \to  \Lambda ^-X}},
\label{gener1}
\end{equation}
where the superscript $+$ or $-$ denotes the helicity.
Within the factorization framework, the production cross sections are described in terms of calculable partonic cross sections and non-perturbative parton distribution and fragmentation functions.
The production cross section has been measured for transverse momenta, $p_\mathrm{T}$, up to about $5\,\mathrm{GeV}/c$ and is well described by perturbative QCD evaluations~\cite{Abelev:2006cs}.
The spin transfer $D_{LL}$ is thus expected to be sensitive to polarized fragmentation function and helicity distribution function of nucleon, as reflected in different model predictions of $D_{LL}$ at RHIC~\cite{deFlorian:1998ba,Boros:2000ex,Ma:2001na,Xu:2002hz}.

The spin transfer $D_{LL}$ in Eq.~(\ref{gener1}) is equal to the polarization of $\Lambda$ $(\bar{\Lambda})$ hyperons ${P}_{\Lambda(\bar\Lambda)}$, if the proton beam is fully polarized.
 ${P}_{\Lambda(\bar\Lambda)}$ can be measured via the weak decay channel $\Lambda \to p \pi^-$ $(\bar \Lambda \to \bar p \pi^+)$ from the angular distribution of the final state,
\begin{equation}
\frac{\mathrm{d}N}{\mathrm{d} \cos{\theta}^*}=\frac{
\sigma\,\mathcal{L}\,A}{2}
(1+\alpha_{\Lambda(\bar\Lambda)} P_{\Lambda(\bar\Lambda)}
\cos{\theta}^*),
\label{ideal}
\end{equation}
where 
$\sigma$ is the $\Lambda$ ($\bar \Lambda$) production cross section and $\mathcal{L}$ is the integrated luminosity. 
$A$ is the detector acceptance, which is in general a function of $\cos\theta^*$ as well as other observables.
$\alpha_{\Lambda}$=$-\alpha_{\bar{\Lambda}} = 0.642 \pm 0.013$~\cite{Amsler:2008zz} is the weak decay parameter, and 
$\theta^*$ is the angle between the $\Lambda$($\bar\Lambda$)  polarization direction and the \mbox{(anti-)proton} momentum in the $\Lambda$ ($\bar\Lambda$) rest frame.

\section{$\Lambda$($\bar\Lambda$) reconstruction at STAR}

In 2005, the Solenoidal Tracker at RHIC (STAR)~\cite{Ackermann:2002ad} collected an integrated luminosity of 2\,$\mathrm{pb}^{-1}$ $pp$ data, with average longitudinal beam polarizations of $52 \pm 3\%$ and $48 \pm 3\%$ for the two beams.
The proton polarization was measured for each beam and each beam fill using Coulomb-Nuclear Interference (CNI) proton-carbon polarimeters~\cite{Jinnouchi:2004up}.
Different beam spin configurations were used for successive beam bunches and the pattern was changed between beam fills.
The data were sorted by beam spin configuration.

The analyzed data sample includes three different triggers. 
One is the minimum bias (MB) trigger sample, defined by a coincidence signal
from Beam-Beam Counters (BBC) on both sides of STAR interaction region.
The other data samples were recorded with the MB trigger condition and with two additional trigger conditions: a high-tower (HT) and a jet-patch (JP).
The HT trigger condition required the proton collision signal in BBC coincidence with a transverse energy deposit $E_\mathrm{T} > 2.6\,\mathrm{GeV}$ in at least one Barrel Electromagnetic Calorimeter (BEMC) tower, covering $\Delta\eta \times \Delta\phi=0.05 \times 0.05$ in pseudorapidity $\eta$ and azimuthal angle $\phi$.
The JP trigger condition imposed the MB condition in coincidence with an energy deposit $E_\mathrm{T} > 6.5\, \mathrm{GeV}$ in at least one of six BEMC patches each covering $\Delta\eta \times \Delta\phi=1 \times 1$.
The total BEMC coverage was $0 < \eta < 1$ with full azimuth in 2005 and became $-1 < \eta < 1$ after 2006.

\begin{figure}
\begin{center}
 \includegraphics[height=.35\textheight]{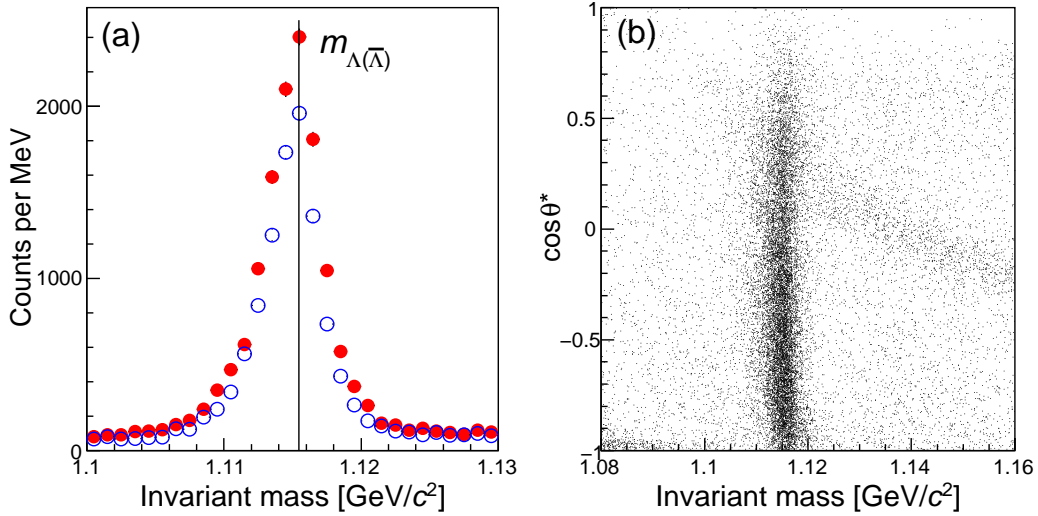}
 \caption{(a) The invariant mass distribution of $\Lambda$ (filled circles) and $\bar \Lambda$ (open circles) candidates from reconstructed $p + \pi^-$ and $\bar p + \pi^+$ track pairs in 2005 MB data after topological selections.
(b) The invariant mass distribution versus $\cos\theta^*$ for $\Lambda$.}
\label{mass}
\end{center}
\end{figure}

The $\Lambda$ ($\bar{\Lambda}$) candidates were identified from the topology of the weak decay channel to $p \pi^-$ ($\bar p \pi^+$), which has a branching ratio of 63.9\%~\cite{Amsler:2008zz}. 
Charged particle tracks in the 0.5\,T magnetic field were measured with the Time Projection Chamber (TPC), covering $0 < \phi < 2\pi$ and $|\eta|< 1.3$.
The charge tracks after particle identification from specific energy loss $dE/dx$ in TPC were paired to form a $\Lambda (\bar{\Lambda})$ candidate and topological selections were applied to reduce background\cite{STAR_DLL,Abelev:2006cs}.
The selections included criteria for the distance of closest approach between the paired tracks and the distance between the point of closest approach and the beam collision vertex, and demanded that the momentum sum of the track pair pointed at the collision vertex.
The criteria were tuned to preserve the signal while reducing the background fraction to 10\% or less.

Figure~\ref{mass}(a) shows the invariant mass distribution for the $\Lambda$ (filled circles) and $\bar\Lambda$ (open circles) candidates reconstructed from MB data with $|\eta| < 1.2$ and $0.3 < p_\mathrm{T} < 3\,\mathrm{GeV}/c$.
The mean values of the $\Lambda$ and $\bar{\Lambda}$ mass distributions are in agreement with the PDG mass value $m_{\Lambda(\bar\Lambda)} = 1.11568\,\mathrm{GeV}/c^2$~\cite{Amsler:2008zz}.
Figure~\ref{mass}(b) shows the same invariant mass distribution versus $\cos\theta^*$ for the $\Lambda$ candidates.
The number of $\Lambda$ candidates varies with $\cos\theta^*$ because of detector acceptance.
The small variation of the reconstructed invariant mass with $\cos\theta^*$ is understood to originate from detector resolution.
In addition to signal, combinatorial background is seen as well as backgrounds of misidentified $e^+e^-$ pairs at low invariant mass values near $\cos\theta^*=-1.0$ and of misidentified $K_{S}^0$ in a diagonal band at high invariant mass values and $\cos\theta^*>-0.2$.

\section{Extraction of $D_{LL}$ and the recent results}

To minimize the uncertainty associated with acceptance, the longitudinal spin transfer $D_{LL}$ was extracted in small $\cos\theta^*$ intervals as follows\cite{STAR_DLL,Xu:2005js}:
\begin{equation}D_{LL}=\frac{1}{\alpha P_\mathrm{beam} \left<\cos \theta^*\right>} \frac{N^+ -R N^-} {N^+ + R N^-},\label{eq_dll}
\end{equation}
where $P_{beam}$ is the beam polarization, $N^+$ ($N^-$) are the $\Lambda$($\bar\Lambda$) counts in a small $\cos\theta^*$ interval when the beam is positively (negatively) polarized, $\left<\cos\theta^*\right>$ is the average value in this $\cos\theta^*$ bin, and $R=L^+/L^-$ is the corresponding luminosity ratio for these two polarization states.
Eq.(3) uses the parity conservation in the hyperon production in $pp$ collisions, which leads to a sign flip of hyperon's longitudinal polarization when the beam helicity is flipped.
The detector acceptance in a $\cos\theta^*$ interval is taken as a constant 
and thus canceled.
For this measurement, only one polarized beam is needed,
and the single spin yields $N^+$ ($N^-$) are formed from the double spin yields $n^{++}$, $n^{+-}$, $n^{-+}$, and $n^{--}$ by beam helicity configuration weighted with their corresponding relative luminosities.
The luminosity ratios were measured with the BBC at STAR \cite{Kiryluk:2005gg}.

 \begin{wrapfigure}[37]{R}{90mm}
 \centering
  \vspace*{-5mm} 
  \includegraphics[width=92mm]{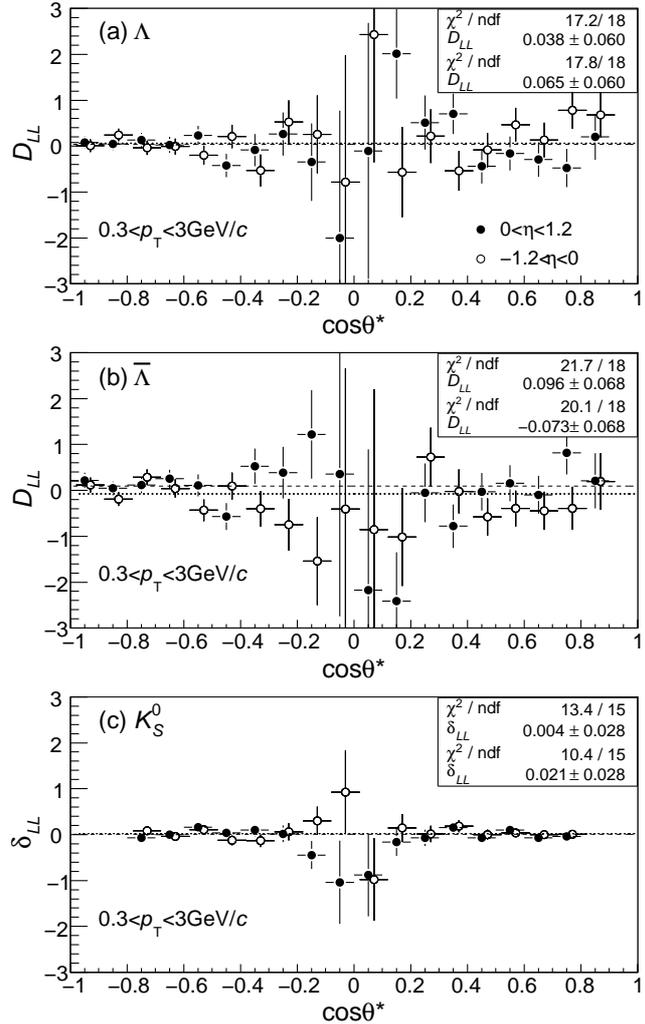}
   \caption{ The spin transfer $D_{LL}$ versus $\cos\theta^*$ for (a) $\Lambda$ and (b) $\bar \Lambda$, and (c) the spin asymmetry $\delta_{LL}$ for the control sample of $K^0_S$ mesons versus $\cos\theta^*$.
Only statistical uncertainties are shown.
The data points with negative $\eta$ have been shifted slightly in $\cos\theta^*$ for clarity.
}
\label{cosin_fit}
\end{wrapfigure}

The yields $N^+$ and $N^-$ were determined for each $\cos\theta^*$ interval from the observed $\Lambda$ and $\bar{\Lambda}$ candidate yields in the mass interval from 1.109 to 1.121\,GeV/$c^2$. 
The corresponding raw values $D^\mathrm{raw}_{LL}$ were averaged over 20 intervals covering the entire $\cos\theta^*$ range.
The obtained $D^\mathrm{raw}_{LL}$ values and their statistical uncertainties were then corrected for (unpolarized) background dilution according to $D_{LL} = D^\mathrm{raw}_{LL}/(1-r)$, where $r$ is the average background fraction.
No significant spin transfer asymmetry was observed for the yields in the sideband mass intervals $1.094 < m < 1.103\,\mathrm{GeV}/c^2$ and $1.127 < m < 1.136\,\mathrm{GeV}/c^2$, and thus no further correction was applied to $D_{LL}$.
However, a contribution was included in the systematic uncertainty of the $D_{LL}$ measurement to account for the possibility that the background could nevertheless be polarized.

The $D_{LL}$ results from the MB data sample versus $\cos\theta^*$ are shown in Fig. \ref{cosin_fit}(a) for $\Lambda$ and Fig.\ref{cosin_fit}(b) for $\bar\Lambda$ hyperons with $0.3 \,< p_\mathrm{T}\, < 3\,\mathrm{GeV}/c$ and $0<\eta<1.2$ and $ -1.2<\eta<0$.
Positive $\eta$ is defined along the direction of the incident polarized beam.
Fewer than 50 counts were observed for $\cos\theta^* > 0.9$ and this interval was discarded for this reason.
The extracted $D_{LL}$ is constant with $\cos\theta^*$, as expected and confirmed by the quality of fit.
In addition, a null-measurement was performed of the spin transfer for the spinless $K_S^0$ meson, which has a similar event topology with a larger production cross section.
The $K_S^0$ candidate yields for $|\cos\theta^*| > 0.8$ were discarded since they have sizable $\Lambda (\bar{\Lambda}$) backgrounds.
The spin transfer result, $\delta_{LL}$, obtained with an artificial weak decay parameter $\alpha_{K_S^0} = 1$, was found consistent with zero, as shown in Fig.~\ref{cosin_fit}(c).

The HT and JP data samples were recorded with trigger conditions that required large energy deposits in the BEMC, in addition to the MB condition.
These triggers, however, did not require a highly energetic $\Lambda$ or $\bar{\Lambda}$.
To minimize the effects of this bias, the HT event sample was restricted to $\Lambda$ or $\bar{\Lambda}$ candidates whose decay (anti-)proton track intersected a BEMC tower that fulfilled the trigger condition.
The $\bar\Lambda$ sample that was selected in this way thus directly triggered the experiment read-out.  
It contains about 1.0$\times 10^4$ $\bar\Lambda$ candidates with $1 < p_\mathrm{T} < 5\,\mathrm{GeV}/c$.

For the JP triggered sample, events were selected with at least one reconstructed jet that pointed to a triggered jet patch.
The same jet reconstruction was used as in Ref.~\cite{Abelev:2006uq}.
The $\Lambda$ and $\bar{\Lambda}$ candidates  whose reconstructed $\eta$ and $\phi$ fell within the jet cone of radius $r_\mathrm{cone} = \sqrt{(\Delta\eta)^2 + (\Delta\phi)^2} = 0.4$ were retained for further analysis.
About $1.3\times10^4~\Lambda$ and $2.1\times10^4~\bar{\Lambda}$ candidates with $1 < p_\mathrm{T} < 5\,\mathrm{GeV}/c$ remain after selections.

\begin{figure}
\begin{center}
 \includegraphics[height=.45\textheight]{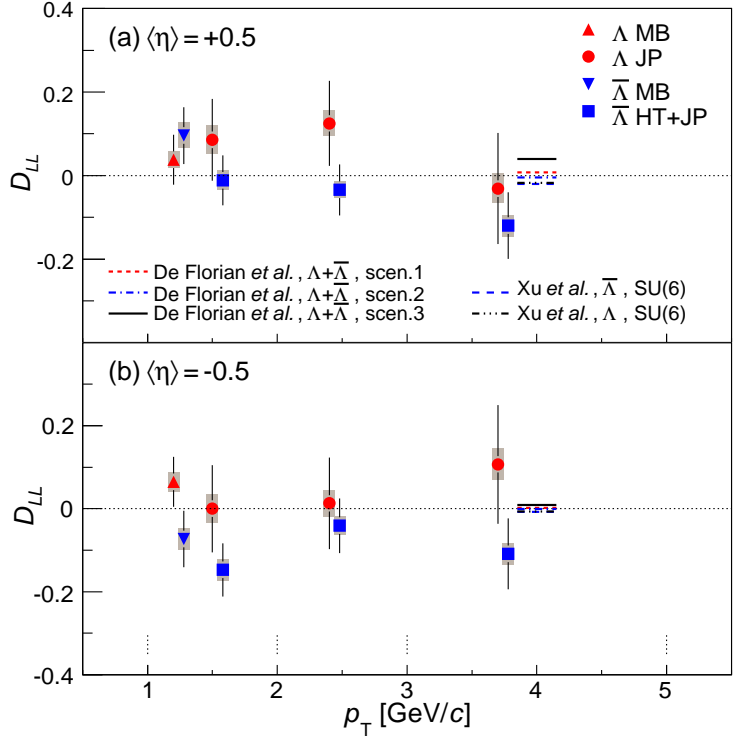}
\caption{ (Color)
Comparison of $\Lambda$ and $\bar \Lambda$ spin transfer $D_{LL}$ in  
polarized proton-proton collisions at $\sqrt{s} = 200$ GeV for (a) positive and  
(b) negative $\eta$ versus $p_\mathrm{T}$.  The vertical bars and bands indicate the  sizes of the statistical and systematic uncertainties, respectively.
The $\bar\Lambda$ data points have been shifted slightly in $p_\mathrm{T}$ for clarity.
The dotted vertical lines indicate the $p_\mathrm{T}$ intervals in the analysis of HT and JP data.
The horizontal lines show model predictions.
}
\label{DLL_05eta}
\end{center}
\end{figure}

Figure~\ref{DLL_05eta} shows the $D_{LL}$ results of $\Lambda$ and $\bar{\Lambda}$ versus $p_\mathrm{T}$ for positive and negative $\eta$ respectively.
The $\bar{\Lambda}$ results from HT and JP data have been combined.
No corrections have been applied for possible decay contributions from heavier baryonic states.
The systematic uncertainty of $D_{LL}$ is varying from 0.02 to 0.04 with increasing $p_\mathrm{T}$.
In estimating the size of the systematic uncertainties, we have combined contributions from the uncertainties in decay parameter $\alpha$ and in the measurements of the proton beam polarization and relative luminosity ratios, as well as uncertainty caused by the aforementioned backgrounds, overlapping events (pile-up), and, in the case of the JP sample, trigger bias studied with Monte Carlo simulation~\cite{STAR_DLL}. 

From Fig.~\ref{DLL_05eta}, the $\Lambda$ and $\bar{\Lambda}$ results for $D_{LL}$ are consistent with each other.
The data have $p_\mathrm{T}$ up to $4\,\mathrm{GeV}/c$, where $D_{LL}= -0.03\pm 0.13(\mathrm{stat}) \pm 0.04(\mathrm{syst})$ for the $\Lambda$ and $D_{LL} = -0.12 \pm 0.08(\mathrm{stat}) \pm 0.03(\mathrm{syst})$ for the $\bar{\Lambda}$ at $\left<\eta\right> = 0.5$.
For reference, the model predictions of Refs.~\cite{deFlorian:1998ba,Xu:2002hz}, evaluated at $\eta=\pm 0.5$ and $p_\mathrm{T}=4\,\mathrm{GeV}/c$, are shown as horizontal lines.
The expectations of Ref.~\cite{deFlorian:1998ba} hold for $\Lambda$ and $\bar{\Lambda}$ combined and examine different polarized fragmentation scenarios, in which the strange (anti-)quark carries all or only part of the $\Lambda$ $(\bar{\Lambda})$ spin.
The model in Ref.~\cite{Xu:2002hz} separates $\Lambda$ from $\bar{\Lambda}$ and otherwise distinguishes the direct production of the $\Lambda$ and $\bar{\Lambda}$ from the (anti-)quark in the hard scattering and the indirect production via decay of heavier (anti-)hyperons.
The evaluations are consistent with the present data and span a range of values that, for positive $\eta$, is similar to the experimental uncertainties.
The values for negative $\eta$ are expected to be negligible and thus less sensitive~\cite{deFlorian:1998ba,Xu:2002hz}.
The current experimental uncertainties are statistics limited.
In 2009, the data sample collected at STAR is about 10 times larger than the above analyzed data sample in 2005, and it is expected that the $p_T$ coverage of $D_{LL}$ measurement be doubled and have the ability to distinguish between different models of polarized fragmentation and parton distribution functions.

In addition to longitudinal spin transfer, the transverse spin transfer of hyperons from the proton is also of particular interest in $pp$ collisions, as it can provide access to the transverse spin content of the nucleon, i.e., the transversity distribution, which is chiral odd and still poorly known in experiment\cite{Barone:2001sp}.
The transverse spin transfer $D_{NN}$ with respect to the production plane
has been measured by the E704 experiment, and
sizable spin transfer was observed at large  $x_F(\equiv 2p_z/{\sqrt s})$ region~\cite{Bravar:1997fb}.
We have started looking at the transverse spin transfer of $\Lambda$ ($\bar\Lambda$) at mid-pseudorapidities with TPC at STAR.
A Forward Hadron Calorimeter (FHC) may be installed behind the available Forward Meson Spectrometer (FMS) in the future at STAR, which may enable the reconstruction of $\Lambda$ hyperons at large $x_F$ region via the decay channel to $n\pi^0$ with $\pi^0$ detected by the FMS and $n$ by the FHC.
The transverse spin transfer measurements in the forward region are expected to be sizable and can provide valuable information for the transversity distribution of nucleon.
In addition, the longitudinal spin transfer $D_{LL}$ of $\Lambda$ ($\bar\Lambda$) in the forward region can provide additional sensitivity to the strange quark polarization\cite{Zhou:2010vm}.

\section{Summary}

In summary, we reported our measurements on the longitudinal spin transfer to $\Lambda$ and $\bar{\Lambda}$ hyperons in $\sqrt{s} = 200\,\mathrm{GeV}$ polarized proton-proton collisions for hyperon $p_\mathrm{T}$ up to $4\,\mathrm{GeV}/c$ with STAR data taken in 2005.
The spin transfer is found to be $D_{LL}= -0.03\pm 0.13(\mathrm{stat}) \pm 0.04(\mathrm{syst})$ for $\Lambda$ and $D_{LL} = -0.12 \pm 0.08(\mathrm{stat}) \pm 0.03(\mathrm{syst})$ for $\bar{\Lambda}$ hyperons with $\left<\eta\right> = 0.5$ and $\left<p_\mathrm{T}\right> = 3.7\,\mathrm{GeV}/c$.
The longitudinal spin transfer is sensitive to the polarized parton distribution and polarized fragmentation functions.
The present results for $\Lambda$ and $\bar{\Lambda}$ have uncertainties that are comparable to the variation between model expectations for the longitudinal spin transfer at RHIC.
With the 2009 data, the $p_T$ coverage and the precision of $D_{LL}$ measurement is expected to be improved significantly.
The possibility of measuring both longitudinal and transverse spin transfer for
$\Lambda$ ($\bar\Lambda$) hyperons in the forward region at STAR is also discussed.

\ack {The author is supported by the Natural Science Foundation of Shandong Province, China, by the Independent Innovation Foundation of Shandong University, and by SRF for ROCS, SEM.}

\end{document}